\documentclass[twocolumn,prl,aps,psfig,showpacs,preprintnumbers,superscriptaddress]{revtex4}
\usepackage{graphicx}
\usepackage{epstopdf}
\usepackage{float}
\usepackage{color}
\usepackage{amsmath}

\begin{document}
\title{Pressure dependence of coherence-incoherence crossover behavior in KFe$_2$As$_2$ observed by
resistivity and $^{75}$As-NMR/NQR}

\author{P.~Wiecki}
\affiliation{Ames Laboratory, U.S. DOE and Department of Physics and Astronomy, Iowa State University, Ames, Iowa  50011  USA}
\author{V.~Taufour$\footnote[1]{Present address: Department of Physics, University of California, Davis, CA 95616, USA}$}
\affiliation{Ames Laboratory, U.S. DOE and Department of Physics and Astronomy, Iowa State University, Ames, Iowa  50011  USA}
\author{D. Y. Chung}
\affiliation{Materials Science Division, Argonne National Laboratory, Argonne, Illinois 60439, USA }
\author{M. G. Kanatzidis}
\affiliation{Materials Science Division, Argonne National Laboratory, Argonne, Illinois 60439, USA }
\affiliation{Department of Chemistry, Northwestern University, Evanston, Illinois 60208, USA}
\author{S.~L.~Bud'ko}
\author{P.~C.~Canfield}
\author{Y.~Furukawa}
\affiliation{Ames Laboratory, U.S. DOE and Department of Physics and Astronomy, Iowa State University, Ames, Iowa  50011  USA}
\date{\today}

\begin{abstract}
    We present the results of $^{75}$As nuclear magnetic resonance (NMR), nuclear quadrupole resonance (NQR), and resistivity measurements in KFe$_2$As$_2$ under pressure ($p$).
    The temperature dependence of the NMR shift, nuclear spin-lattice relaxation time ($T_1$) and resistivity show a crossover  between a high-temperature incoherent, local-moment behavior and a low-temperature coherent behavior at a crossover temperature ($T^*$). 
     $T^*$ is found to increase monotonically with pressure, consistent with increasing hybridization between localized $3d$ orbital-derived bands with the itinerant electron bands. 
    No anomaly in $T^*$ is seen at the critical pressure $p_{\rm c}=1.8$ GPa where a change of slope of  the superconducting (SC) transition temperature $T_{\rm c}(p)$ has been observed. 
   In contrast, $T_{\rm c}(p)$  seems to correlate with antiferromagnetic spin fluctuations in the normal state as measured by the NQR  $1/T_1$ data, although such a correlation cannot be seen in the replacement effects of A in the AFe$_2$As$_2$ (A= K, Rb, Cs) family.
    In the superconducting state, two $T_1$  components are observed at low temperatures, suggesting the existence of two distinct local electronic environments. 
    The temperature dependence of the short $T_{\rm 1s}$ indicates nearly gapless state below $T_{\rm c}$. 
    On the other hand, the temperature dependence of the long component 1/$T_{\rm 1L}$ implies a large reduction in the density of states at the Fermi level due to the SC gap formation.  
    These results suggest a real-space modulation of the local SC gap structure in KFe$_2$As$_2$ under pressure.

\end{abstract}

\maketitle

\section{I. Introduction}

    The iron-based superconductors (SCs) continue to be the focus of intense research in condensed matter physics, due to their unique interplay of magnetic, orbital and charge degrees of freedom \cite{Canfield2010,Johnston2010,Fernandes2014,Si2016,Dai2015}. 
   Among the iron-based SCs, the heavily hole-doped iron-pnictide superconductor KFe$_2$As$_2$, with a SC transition temperature of $T_{\rm c}\sim3.5$ K, shows several unique properties.
    The Sommerfeld coefficient ($\gamma\sim102$ mJ/molK$^2$) is significantly enhanced, 
    and the magnetic susceptibility exhibits a broad peak around 100 K \cite{Hardy2013}. 
Nuclear magnetic resonance (NMR) spin-lattice relaxation rates ($1/T_1$) are strongly enhanced, evidencing antiferromagnetic spin fluctuations. 
    Curie-Weiss fits to the NMR data have demonstrated the proximity of KFe$_2$As$_2$ to a quantum critical point (QCP)  \cite{Hirano2012,Wiecki2015,Wang2016}.
     These results indicate a heavy quasiparticle effective mass and strong electronic correlations \cite{Hardy2013,Wiecki2015}.
     Recent NMR investigations have also pointed out the importance of ferromagnetic spin correlations in this material \cite{Wiecki2015}.

    Furthermore, the SC properties of KFe$_2$As$_2$ are also unique. 
    Whereas two full SC gaps are reported in the hole-doped series Ba$_{1-x}$K$_x$Fe$_2$As$_2$ for $x$ $<$ $\sim$0.8 \cite{Ding2008}, a nodal SC gap structure 
    in KFe$_2$As$_2$ ($x=1$) has been suggested by several experiments \cite{Fukazawa2009,Hashimoto2010,KawanoFurukawa2011,Reid2012,Okazaki2012,Ota2014}.   
    A large full gap accompanied by several very small gaps has also been proposed based on specific heat measurements \cite{Hardy2014}. 
    In addition, $T_{\rm c}$ shows non-monotonic behavior under pressure, with a minimum at $p_{\rm c}\sim1.8$ GPa, which has been suggested to be caused by a change in the SC gap structure \cite{Tafti2013,Terashima2014,Taufour2014}. 
    Measurements of the pressure dependence of the upper critical field $H_{c2}$ suggested the appearance of a $k_z$ modulation of the SC gap above $p_{\rm c}$ \cite{Taufour2014}.

     Analogous behavior has also been found in the related alkali metal compounds RbFe$_2$As$_2$ and  CsFe$_2$As$_2$ \cite{Hong2013,Tafti2014,Zhang2015,Tafti2015,Mizukami2016,Eilers2016,Civardi2016}, which show even greater mass enhancements with $\gamma\sim127$ mJ/molK$^2$ and $\gamma\sim184$ mJ/molK$^2$, respectively \cite{Wu2016}.
   The unusual properties of the AFe$_2$As$_2$ (A = K, Rb, Cs) family have been pointed out \cite{Hardy2013,Wu2016,Zhao2017} to be quite similar to 
$f$-electron heavy fermion materials \cite{Gegenwart2008,Curro2009}, which display a crossover between a high-temperature incoherent, local-moment behavior and a low-temperature coherent behavior, with the crossover occurring at a temperature $T^*$.
      In this picture, the importance of dual role of Fe $d$ electrons has been pointed out theoretically\cite{You2011, Gorkov2013} where the two aspects of the itinerant and localized electrons may originate from different 3$d$ orbitals of the iron ions. 
   Recently, experimental \cite{Fang2015,Zhao2017} and theoretical \cite{Hardy2013} studies suggest that the bands derived from the Fe 3$d_{xy}$ orbitals would play the role of the local moments. 
    This orbital-selective localization is due to the strong Hund coupling in these materials \cite{Georges2013}.

     Recent NMR measurements have pointed out a possible $d$-electron heavy fermion behavior in the AFe$_2$As$_2$ (A = K, Rb, Cs) family at ambient pressure \cite{Wu2016}.
     $T^*$ is reported to increase from 85 K for Cs, to 125 K for Rb and to 165 K for KFe$_2$As$_2$. 
    Thermal expansion measurements on this family also find the lowest $T^*$ for Cs and highest $T^*$ for K, although the reported crossover temperatures are lower \cite{Hardy2016}.
    Since the so-called chemical pressure effects would increase when one moves from Cs to Rb  to K due to the decrease in size of the alkali metal ion, this suggests that $T^*$  increases with increasing the chemical pressure. 
    Furthermore, two empirical relationships involving $T^*$ have been discussed \cite{Wu2016}. 
    First, the superconducting transition temperature $T_{\rm c}$ is generally proportional to $T^*$, that is $T_{\rm c} \propto T^*$, reflecting the  correlation of $T_{\rm c}$ to local magnetic coupling $J$ as pointed out in Ref. \onlinecite{Pines2013} in the context  of $f$-electron heavy fermion SCs. 
   Second, the Sommerfeld coefficient $\gamma$, and thus the effective mass $m^*$, is inversely proportional to $T^*$, that is $\gamma^{-1}\propto T^*$ (see also Ref. \onlinecite{Hardy2013}).

     The $T_{\rm c} \propto T^*$ relationship for the  AFe$_2$As$_2$ (A = K, Rb, Cs) naively suggests that the non-monotonic behavior of $T_{\rm c}$ in these materials under pressure could be due to a non-monotonic behavior of $T^*$ under pressure.
     This motivates an experimental investigation of the relationship between $T_{\rm c}$ and $T^*$ under  pressure. 
   Here, we have carried out NMR and nuclear quadrupole resonance (NQR) measurements under high pressure up to 2.1 GPa and resistivity measurements up to $\sim5$ GPa in order to investigate the pressure dependence of $T^*$ and to test its relationship with $T_{\rm c}$.  
     Based on the NMR and resistivity data, we find that $T^*$  increases monotonically with increasing pressure with no anomaly associated with crossing $p_{\rm c}\sim1.8$ GPa.
   These results indicate that $T^*$ is not the primary driver of the pressure dependence of $T_{\rm c}$ in KFe$_2$As$_2$. 
   On the other hand, $1/T_1$ measurements demonstrate that spin fluctuations are suppressed with increasing pressure up to the $p_{\rm c}$ and then start to be enhanced above $p_{\rm c}$, suggesting that $T_{\rm c}$ is related to spin fluctuations in the normal state. 
    In the superconducting state, two-component NQR relaxation is observed below $T=1$ K, suggesting real space variation of the superconducting gap structure. 
    One of the two components, the short $T_1$ component,  shows no change in the slope of $1/T_1$ across $T_{\rm c}$ above 1.5 GPa, indicating these nuclei see a gapless local electronic environment in the SC state under these pressure conditions. 
   Only the second component, the long  $T_1$ component,   shows a large reduction of the density of states at the Fermi energy due to the SC gap.

\section{II. Experimental Details}

   Highly pure  KFe$_2$As$_2$ crystal sample was obtained by recrystallization of pre-reacted  KFe$_2$As$_2$ polycrystalline powder in KAs flux as follows. 
    KFe$_2$As$_2$ polycrystalline powder was prepared by annealing a stoichiometric mixture of K/Fe/As (0.27/0.77/1.03 g) contained in an alumina crucible which was subsequently sealed in a sealed silica tube under vacuum, at 700 $^{\circ}$C for one day. 
   KAs was prepared by heating a stoichiometric 1/1 ratio of K/As (0.94/1.80 g) in an alumina tube sealed in a silica tube at 250 $^{\circ}$C for 12 h. 
   The obtained KFe$_2$As$_2$ powder was then thoroughly mixed with KAs at a ratio of 1/4 (1.10/1.67 g) and heated to 1,050 $^{\circ}$C for 12 h, followed by cooling slowly to room temperature at 5 $^{\circ}$C/h. 
     Isolation of KFe$_2$As$_2$ crystals from excess KAs flux was performed by dissolving KAs in ethanol for two days under nitrogen gas flow, which produces very shiny thin plate KFe$_2$As$_2$ crystals. 
    The quality of KFe$_2$As$_2$ crystals was confirmed by a very sharp superconducting transition at 3.4 K from the magnetic susceptibility measurement.

     $^{75}$As-NMR/NQR ($I=3/2$; $\gamma/2\pi=7.2919$ MHz/T; $Q=0.29$ barns) measurements  were performed using a lab-built, phase-coherent, spin-echo pulse spectrometer. 
    The KFe$_2$As$_2$ sample was a fine powder in order to maximize the surface area for NMR/NQR measurements. 
    The total mass of powder used in the high-pressure NMR/NQR measurements was $\sim 15$ mg. 
    The $^{75}$As-NMR spectra were obtained either by sweeping the magnetic field $H$ at a fixed frequency $f$ = 54.8756 MHz or by Fourier transform of the NMR echo signals at a constant magnetic field of $H$ = 7.41 T.
    $^{75}$As-NQR spectrum in zero field was measured in steps of frequency by  measuring the intensity of the Hahn spin-echo.  
    For our measurements at ultra-low temperatures below 1 K, we used a dilution refrigerator (Oxford Instruments, Kelvinox 100) where the pressure cell was mounted. 

    The $^{75}$As NMR/NQR $1/T_1$ was measured with a recovery method using a single $\pi$/2 saturation pulse. For NMR measurements, the $1/T_1$ at each $T$ was determined by fitting the nuclear magnetization $m$ versus time $t$ using the exponential function
\begin{equation}
1-\frac{m(t)}{m(\infty)}=0.1\exp{(-t/T_1)}+0.9\exp{(-6t/T_1)},
 \end{equation}
 where $m(t)$ and $m(\infty)$ are the nuclear magnetization at time $t$ after the saturation and the equilibrium nuclear magnetization at $t \to \infty$, respectively.
For NQR measurements, the recovery curve was fit to 
\begin{equation}
1-\frac{m(t)}{m(\infty)}=\exp{(-3t/T_1)}.
\end{equation}

Pressure was applied at room temperature using a hybrid CuBe/NiCrAl piston-cylinder-type high pressure clamp cell \cite{Fujiwara2007,Aso2007}. 
Daphne 7373 was chosen as the pressure transmitting medium. 
Pressure calibration was accomplished by $^{63}$Cu-NQR in Cu$_2$O \cite{Fukazawa2007,Reyes1992} at 77 K. 
In our pressure cell, the sample pressure decreases by $\sim$0.2 GPa when cooled from room temperature to 100 K, but remains constant below 100 K. 
The NMR coils inside the pressure cell consisted of $\sim$20 turns of 40AWG copper wire. 
   The sample and calibration coils were oriented with their axes perpendicular to each other to avoid interference between coils. 
   
   The single-crystal electrical resistivity measurements were performed using the four-probe method
   with current in the $ab$ plane \cite{Taufour2014}. Pressure was applied at room temperature using a modified Bridgman cell \cite{Colombier2007}
   with a 1:1 mixture of n-pentane:isopentane as a pressure medium, with the pressure determined using the superconducting 
   transition of Pb.
   

\section{III. Results and Discussion}

\subsection{A.  $T_{\rm c}$ and critical pressure}
\begin{figure}[t]
\centering
\includegraphics[width=\columnwidth]{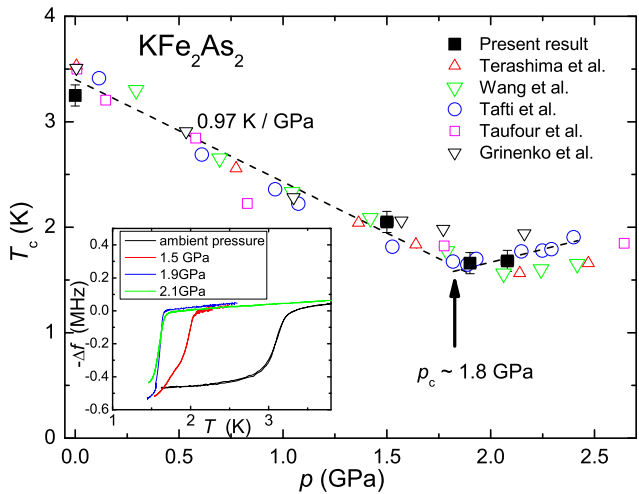}

\caption{
Superconducting transition temperature $T_{\rm c}$ as a function of pressure determined by onset of Meissner effect measured  by {\it in situ} ac-susceptibility.
$p_{\rm c}\sim1.8$  GPa marks the critical pressure where $T_{\rm c}$ changes slope. Previously reported data are shown for comparison:
Terashima {\it et al}. Ref. \onlinecite{Terashima2014}; Wang {\it et al}. Ref. \onlinecite{Wang2016}; 
Tafti {\it et al}. Ref. \onlinecite{Tafti2013}; Taufour {\it et al}. Ref. \onlinecite{Taufour2014}; Grinenko {\it et al}. Ref. \onlinecite{Grinenko2014}.
Inset shows the typical temperature dependence of the change in the NMR coil tank circuit resonance frequency, $\Delta f$, under different pressures.
}
\label{fig:Tc}
\end{figure}

     The superconducting transition temperature $T_{\rm c}$ of the KFe$_2$As$_2$ powder was determined by measuring the $T$ dependence of the NMR coil tank circuit resonance frequency, $f(T)$, under zero magnetic field. 
    The frequency $f$ is a measure of the ac-susceptibility $\chi_{\rm{ac}}(\omega_{\rm{NMR}})$  since $f=1/2\pi\sqrt{LC}$ and $L=L_0(1+\chi_{\rm{ac}})$. The onset of the Meissner effect therefore results in a sharp change of $f(T)$ as shown in the inset of Fig. \ref{fig:Tc}.
    At ambient pressure, we find $T_{\rm c}\sim3.3$ K, as expected. 
    The pressure dependence of $T_{\rm c}$ is shown in Fig. \ref{fig:Tc} together with the data reported previously \cite{Tafti2013,Taufour2014,Terashima2014,Wang2016,Grinenko2014,Budko2012}. 
     $T_{\rm c}$ decreases with $p$ below the critical pressure $p_{\rm c}\sim1.8$ GPa with a rate of 0.97 K/GPa, while $T_{\rm c} $ shows weak pressure dependence above $p_{\rm c}$. 


\subsection{B.  NMR spectrum}

\begin{figure}[t]
\centering
\includegraphics[width=\columnwidth]{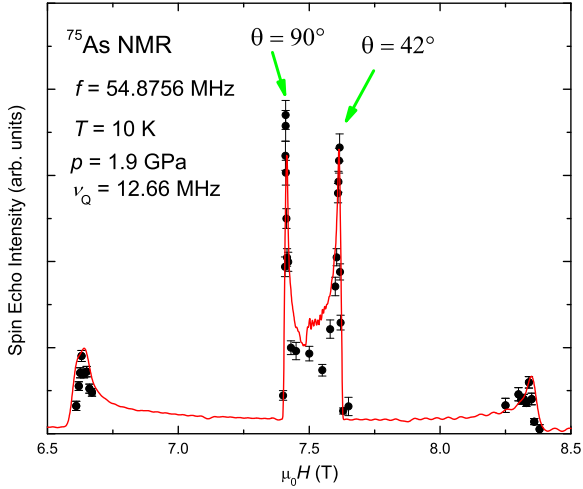}

\caption{
Representative field-swept $^{75}$As NMR spectrum of KFe$_2$As$_2$ powder measured at $T$ = 10 K and $p$ = 1.9 GPa. 
The central transition line is split into two lines by the second order quadrupole effect. 
$\theta$ is the angle between the external field and the principal axis of the electric
field gradient (see text).  The red curve is a simulated powder spectrum with $\nu_Q=12.66$ MHz.
}
\label{fig:NMRspect}
\end{figure}

Figure \ref{fig:NMRspect} shows a representative field-swept NMR spectrum of the KFe$_2$As$_2$ powder measured at 10 K and $p$ = 1.9 GPa.
The spectrum is typical for an $I=3/2$ nucleus in a powder sample with Zeeman interaction greater than quadrupole interaction. 
A central transition is flanked by two satellite lines split by the quadrupole interaction of the As nucleus with the local electric field gradient (EFG). 
   In addition, the central transition line is split by the second-order quadrupole perturbation.

The situation is described by the spin Hamiltonian \cite{Slichter_book} 
\begin{equation}
{\cal H}=-h\nu_L(1+K_{z'})I_{z'}+\frac{h\nu_Q}{6}(3I_z^2-I^2),
\end{equation}
appropriate for tetragonal crystals.
Here $z'$ is the direction of the applied field ($H_{\rm ext}$) and $z$ is the direction of the principal axis of the EFG.
    $\nu_L=\gamma H_{\rm ext}/2\pi$ is the Larmor frequency and $K_{z'}$ represents the NMR shift. 
    The quadrupole frequency for an $I=3/2$ nucleus can be expressed as $\nu_Q=e^2QV_{zz}/2h$, $e$ is the electron charge, $Q$ is the nuclear quadrupole moment, $V_{zz}$ is the EFG and $h$ is Planck's constant. 
     According to this Hamiltonian, the NMR spectrum  depends on the angle $\theta$ between the external field and the EFG principal axis.
To first order, the quadrupole satellite resonance frequencies are given by 
\begin{equation}
\nu_\pm=\nu_L(1+K_{z'})\pm\frac{\nu_Q}{2}\left(3\cos^2\theta-1\right)
\end{equation}
In second order perturbation theory, the central transition frequency depends on $\theta$ according to
\begin{equation}
\nu(\theta)=\nu_L\left(1+K_{z'}\right)
-\frac{3\nu_Q^2}{16\nu_L}\sin^2\theta(9\cos^2\theta-1).
\label{eq:2ndorder}
\end{equation}

    In a powder sample, crystallites with all values of $\theta$ are present.
    Under these conditions the quadrupole satellites appear as sharp peaks at $\nu_L(1+K_{z'})\pm\nu_Q/2$ which correspond to $\theta=90^{\circ}$.
For a powder, sharp peaks are observed in the central transition for $\theta=90^{\circ}$ and $\theta=\cos^{-1}(\sqrt{5/9})=41.8^{\circ}$, as shown by the calculated powder-pattern spectrum in Fig. \ref{fig:NMRspect}.
   The calculated spectrum assumes no preferential orientation of crystal grains, which is reasonable because the solidifications of the pressure medium prevent the crystal grains from re-orienting.  
   In a field-swept spectrum, the $\theta=90^{\circ}$ peak occurs at lower field, as indicated in Fig. \ref{fig:NMRspect}.
    Since the EFG principal axis is along the $c$ direction in KFe$_2$As$_2$, the $\theta=90^{\circ}$ peak arises from 
those crystallites that experience an external field in the crystal $ab$ plane. 
    We conducted our NMR shift and $1/T_1$ measurements at this peak of the central transition. 

\begin{figure}[t]
\centering
\includegraphics[width=\columnwidth]{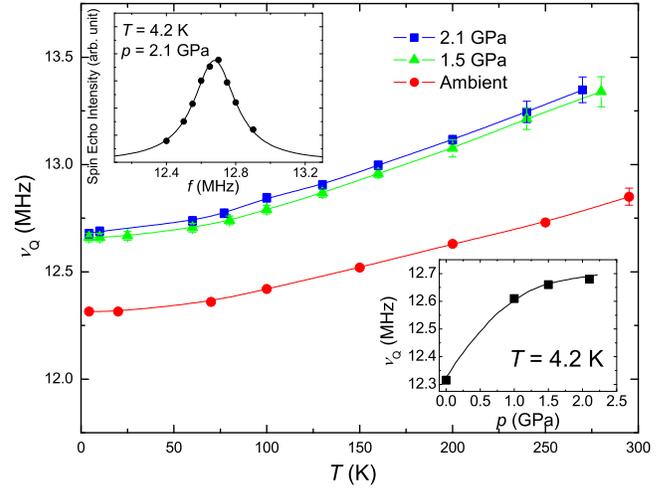}

\caption{
Nuclear quadrupole resonance (NQR)  frequency as a function of temperature for indicated pressures. 
Upper inset: Representative NQR spectrum at $p$ = 2.1 GPa and $T$ = 4.2 K, shown with a Lorentzian fit.
Lower inset: NQR frequency as a function of pressure at $T$ = 4.2 K. Lines are guides to the eye. 
}
\label{fig:fQ}
\end{figure}

The quadrupole resonance frequency $\nu_Q$ was obtained by a direct measurement of the NQR spectrum at zero magnetic field.
The typical NQR spectrum is shown in the inset of Fig. \ref{fig:fQ}, where the full-width-at-half-maximum (FWHM) of the NQR spectrum is $\sim 250$ kHz at $T=4.2$ K, 
which is consistent with the value reported previously \cite{Wang2016} and is sharper than early NQR data measured at ambient pressure \cite{Fukazawa2009}.
The temperature and pressure dependence of $\nu_Q$ is summarized in Fig. \ref{fig:fQ}.
As a function of temperature, $\nu_Q$ is nearly constant below 50 K, and increases slowly above 50 K, which is not simply explained by the so-called $T^{3/2}$-law  originating from the thermal vibrations of the lattice \cite{Mossbaure}.  
A similar increase of $\nu_Q$ at the Fe site is observed by M\"ossbauer measurements \cite{Budko2017}.
 It is interesting to note that the value and temperature dependence of $\nu_Q$ in KFe$_2$As$_2$ is very 
similar to the $\nu_Q$ measured at the As(1) site near the K layer in the recently discovered superconductor CaKFe$_4$As$_4$, where magnetic fluctuations are involved to explain the temperature dependence \cite{Cui2017}. 
As a function of pressure at constant temperature, $\nu_Q$ increases quickly up to 1.5 GPa, but increases slowly thereafter as in seen in the inset to Fig. \ref{fig:fQ}.  
Similar pressure dependence of $\nu_Q$ in KFe$_2$As$_2$ has been reported \cite{Wang2016}.
No sharp anomalies are seen in $\nu_Q$, indicating no  structural phase transitions in the measured pressure and temperature range. 

    In order to precisely determine the NMR shift with external field applied in the $ab$ plane,  we performed Fourier transform measurements of the $\theta=90^{\circ}$ peak of the NMR central transition line at a constant magnetic field.
In general,  the central transition frequency is given by 
\begin{align}
\nu(\theta)&=\nu_L\left(1+\frac{2K_{\rm ab}+K_{\rm c}}{3}\right)\nonumber \\
&\qquad {}+\frac{\nu_L}{3}(K_{\rm c}-K_{\rm ab})(3\cos^2\theta-1)\nonumber \\
&\qquad {}-\frac{3\nu_Q^2}{16\nu_L}\sin^2\theta(9\cos^2\theta-1).
\label{eq:2ndorderaniso}
\end{align}
where $K_{\rm ab}$ and $K_{\rm c}$ are Knight shifts for $H$ $\parallel$ $ab$ plane and $H$ $\parallel$ $c$ axis, respectively. 
   In the present case, since $\frac{3\nu_Q^2}{16\nu_L}\gg\frac{\nu_L}{3}(K_{\rm c}-K_{\rm ab})$ (Ref. \onlinecite{Hirano2012} gives $|K_{\rm c}-K_{\rm ab}|\sim0.001$),  Eq. \ref{eq:2ndorderaniso} can be simplified as 
\begin{equation}
\nu(\theta=90^{\circ})=\nu_L(1+K_{\rm ab})+\frac{3\nu_Q^2}{16\nu_L}
\end{equation}
when $\theta=90^{\circ}$. 
    We therefore obtain $K_{\rm ab}$ by subtracting $3\nu_Q^2/16\nu_L$ from the measured resonance frequency, $\nu(\theta=90^{\circ})$.

The obtained NMR shifts are shown in Fig. \ref{fig:Kab}.
At ambient pressure, the NMR shift is nearly constant at low temperature and shows a broad peak near 150 K, before decreasing at high temperature. 
The behavior of $K_{\rm ab}$ is qualitatively similar under pressure, with the broad peak shifting to slightly higher temperature.

\subsection{C.  Crossover temperature $T^*$}

\begin{figure}[t]
\centering
\includegraphics[width=\columnwidth]{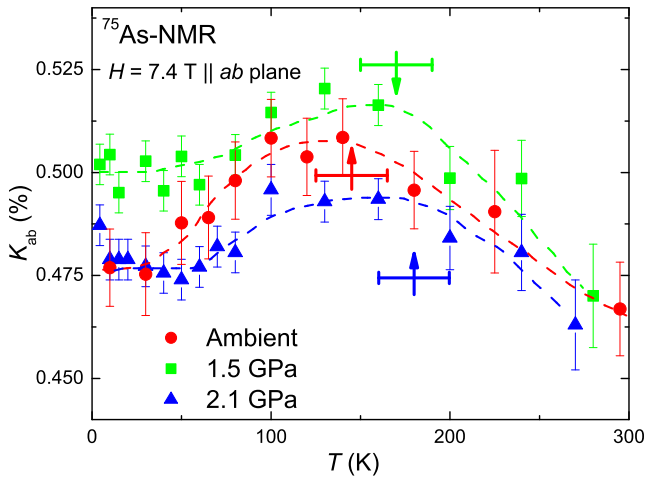}
\caption{
NMR shift with external field aligned in the $ab$ plane ($ K_{\rm ab}$) for indicated pressures. 
The dashed lines are guides to the eye.
The arrows represent the  crossover temperature $T^*_{\rm NMR}$
as determined by NMR $1/T_1$ measurements (see text and Fig. \ref{fig:T1NMR}).
The horizontal bars denote the uncertainty in estimation of $T^*_{\rm NMR}$ ($\pm20$ K).
}
\label{fig:Kab}
\end{figure}

The NMR shift data in Fig. \ref{fig:Kab} are consistent with a coherence/incoherence crossover behavior in KFe$_2$As$_2$  at all measured pressures. 
The broad peak in the NMR shift has been interpreted as  the crossover from the high-temperature local-moment (Curie Weiss) behavior
to the low temperature coherent state \cite{Hardy2013,Wu2016}. 
We could not reliably extract the crossover temperature $T^*$ from the NMR shift data alone 
because of the weak temperature dependence of the NMR shift and also the broad quadrupole powder lineshape, although the data suggest a small increase of $T^*$ under pressure.

\begin{figure}[tb]
\centering
\includegraphics[width=\columnwidth]{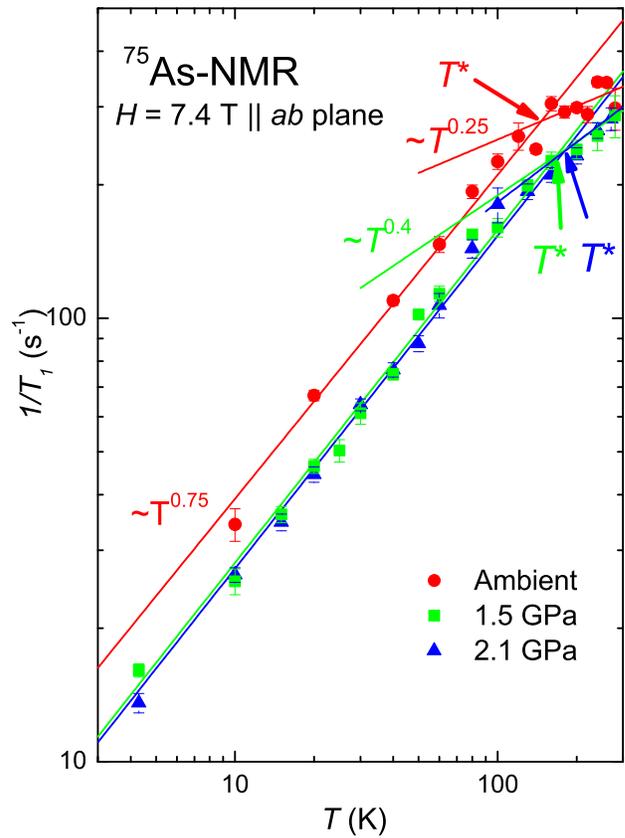}

\caption{
NMR spin-lattice relaxation rate $1/T_1$ as a function of temperature.
The coherence/incoherence crossover temperature $T^*_{\rm NMR}$ is found by the change of slope of $1/T_1$.
The uncertainty in estimation of $T^*_{\rm NMR}$ is $\pm20$ K (see text).
}
\label{fig:T1NMR}
\end{figure}

    The coherence/incoherence crossover temperature in KFe$_2$As$_2$ can also be estimated from the nuclear spin-lattice relaxation rate $1/T_1$ data, shown in Fig. \ref{fig:T1NMR}.
   Our results for $1/T_1$ at ambient pressure are quantitatively consistent with Ref. \onlinecite{Hirano2012}.
    At low temperature, $1/T_1$ shows a power law behavior  $1/T_1\sim T^{0.75}$ for all pressures, as seen in Fig. \ref{fig:T1NMR}. 
   An obvious reduction in the slope of $1/T_1$ is seen at high temperature, however. 
    Similar temperature dependence of 1/$T_1$ is often observed in heavy fermion systems, where 1/$T_1$ shows a power law behavior of 1/$T_1$ $\propto$ $T^\alpha$  (i.e. $\alpha=0.25$ in CeCoIn$_5$ \cite{Kohori2001} and $\alpha=1$ in URu$_2$Si$_2$ \cite{Kohara1986}) at low temperatures due to coherent metallic heavy fermion states and levels off at higher temperatures due to incoherent local moment behaviors.
    Thus the change in slope of the temperature dependence of  1/$T_1$ gives an estimate of the coherence/incoherence crossover temperature (defined as $T^*_{\rm NMR}$). 
    From the $T_1$ data, we find $T^*_{\rm NMR}\sim145\pm20$ K at ambient pressure, $T^*_{\rm NMR}\sim170\pm20$ K at 1.5 GPa and $T^*_{\rm NMR}\sim180\pm20$ K at 2.1 GPa, indicating that $T^*_{\rm NMR}$ increases under pressure. 
The uncertainty in $T^*_{\rm NMR}$ is due primarily to uncertainty in the high-$T$ slope (see below).
   These values of $T^*_{\rm NMR}$ seem to be  consistent with the high-temperature end of the broad peak of $K_{\rm ab}$ (arrows in Fig. \ref{fig:Kab}).
   The increase of $T^*_{\rm NMR}$ under pressure is reasonable, as the application of pressure should increase the hybridization between localized  and itinerant electrons, thus increasing the local magnetic coupling $J$ \cite{Pines2013}.
      
      We also note that 1/$T_1$ constant behavior above the coherent/incoherent crossover temperature $T^*$ is observed in  CsFe$_2$As$_2$ \cite{Wu2016}, which has the highest effective mass of the AFe$_2$As$_2$ (A = K, Rb, Cs) family and therefore most localized electrons. 
    However, as seen in Fig. \ref{fig:T1NMR}, in KFe$_2$As$_2$ at ambient pressure $1/T_1$ does not level off completely above $T^*_{\rm NMR}$  but rather increases much more slowly, following roughly $1/T_1\sim T^{0.25\pm0.1}$. 
   Furthermore, as $T^*_{\rm NMR}$ increases under pressure, so does the slope of $1/T_1\sim T^{0.4\pm0.1}$. 
   It would be interesting if the high-temperature slope correlates with extent of the localization.

\begin{figure}[tb]
\centering
\includegraphics[width=\columnwidth]{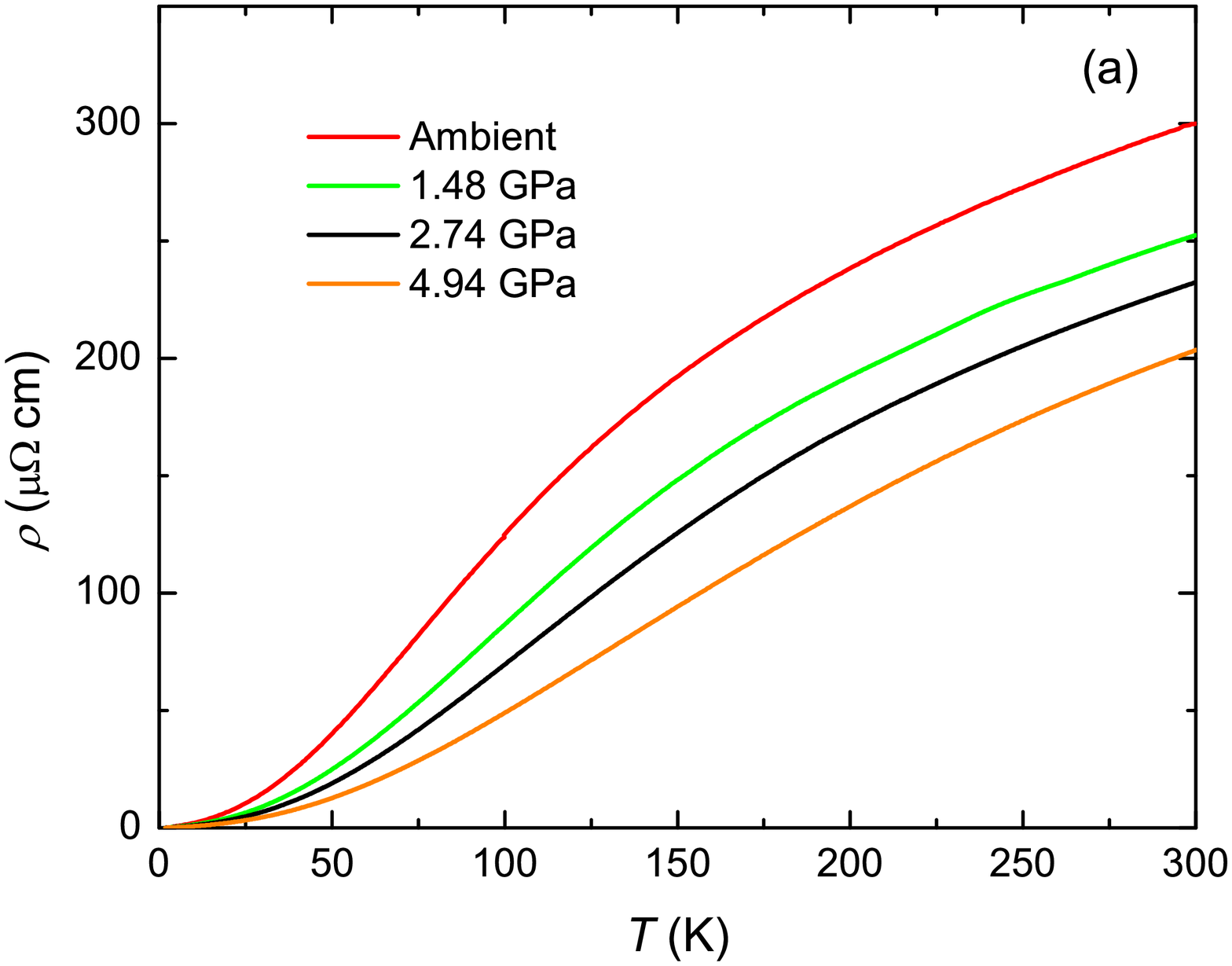}
\includegraphics[width=\columnwidth]{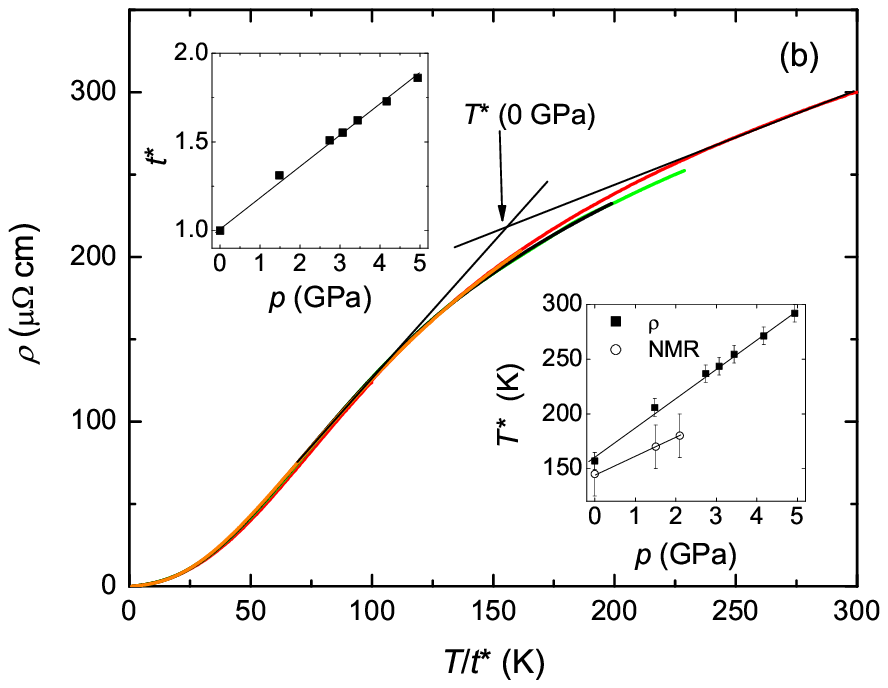}
\caption{
(a) Resistivity of KFe$_2$As$_2$ single crystals \cite{Taufour2014} for selected pressures. 
(b) Resistivity plotted as a function of $T/t^*$ where the scaling coefficient $t^*$ is chosen so as to merge each curve with the ambient pressure curve. 
For ambient pressure, $t^*\equiv1$.
Upper inset: pressure dependence of the unitless scaling factor $t^*$. 
Lower inset: Comparison of pressure dependence of $T^*$ as measured by resistivity ($T^*_{\rm R}$; filled symbols) and NMR ($T^*_{\rm NMR}$; open symbols). 
For resistivity $T^*_{\rm R}$ $ =$ (157 K)$t^*$, where
$T^*_{\rm R}$  at $0$ GPa is determined by the crossing of two tangent lines, as proposed in Ref. \onlinecite{Wu2016} (see text).
}
\label{fig:rho}
\end{figure}

    To corroborate our estimate of $T^*$  and expand the results to pressures higher that those attainable in our NMR pressure cell, we also present  and re-analyze single-crystal resistivity data up to $\sim5$ GPa \cite{Taufour2014}, as shown in Fig. \ref{fig:rho}(a). 
    In heavy fermion systems, one expects a decrease of the resistivity below the coherence temperature, often showing a broad maximum at $T^*$ \cite{Yang2008,Hardy2013}. 
     While the NMR data provide incontrovertible evidence for coherence-incoherence crossover, the resistivity contains contributions from phonon scattering which complicate the interpretation of the data. 
    The decrease in resistivity observed in Fig. \ref{fig:rho}(a) could, in principle, be due to the small Debye temperature and not electronic coherence effects. 
    However, in the AFe$_2$As$_2$ (A = K, Rb, Cs) family, a strong correlation has been observed between $T^*_{\rm NMR}$ (as observed by NMR) and the cross point of two approximately linear trends in the resistivity \cite{Wu2016}. 
    This method, then, appears to give a reasonable estimate of $T^*$ in these materials.
     Here we also apply this method to estimate  $T^*$ (defined as $T^*_{\rm R}$) in KFe$_2$As$_2$ using the resistivity data.
     We note that our resistivity curves for different pressures can be scaled by a pressure dependent scaling factor $t^*$ (defined dimensionless), as shown in Fig. \ref{fig:rho}(b). 
    The pressure dependence of $t^*$ is shown in the upper inset. 
    To estimate $T^*_{\rm R}$  from the resistivity data, we use the cross point of two approximately linear trends as shown in Fig. \ref{fig:rho}(b) where  $T^*_{\rm R}$ is estimated to be  $T^*_{\rm R} =157$ K for the ambient pressure data. 
     Then, the pressure dependence of  $T^*_{\rm R}$ can be obtained by using the pressure dependence of $t^*$.
     As shown  in the lower inset of Fig. \ref{fig:rho}(b),  $T^*_{\rm R}$ increases with increasing pressure.
     While the values of $T^*_{\rm R}$ extracted from the resistivity data up to 2.1 GPa are slightly higher than the $T^*_{\rm NMR}$ values identified by NMR data,
   both techniques show the increase of the coherent/incoherent crossover temperature $T^*$ with applied pressure.
    It is clear that $T^*_{\rm R}$ evolves continuously, showing no anomaly at $p_{\rm c}\sim1.8$ GPa. 
     It is interesting to note that  the resistivity data for the Rb- and Cs-samples \cite{Wu2016} can also be scaled to our ambient pressure data with $t^*=0.78$ ($T^*_{\rm R}=123$ K) and $t^*=0.52$ ($T^*_{\rm R}=82$ K) respectively.

\begin{figure}[t]
\centering
\includegraphics[width=\columnwidth]{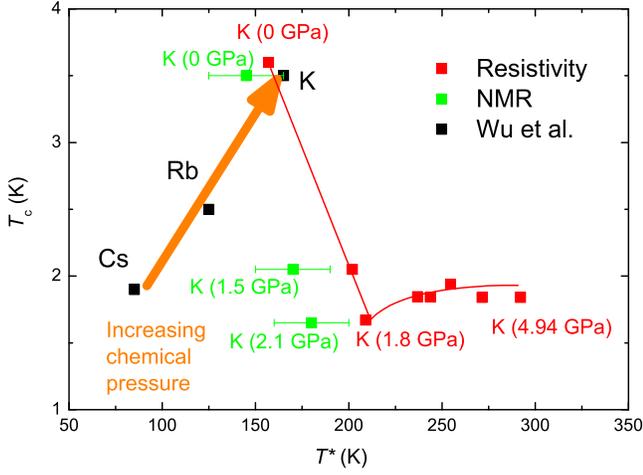}

\caption{
Plot of $T_{\rm c}$ vs  $T^*$ for AFe$_2$As$_2$ (A = K, Rb, Cs) family at ambient pressure \cite{Wu2016} (black). 
The orange arrow illustrates the increase of chemical pressure from CsFe$_2$As$_2$ to KFe$_2$As$_2$.
The green data plots $T_{\rm c}$ vs $T^*$ for KFe$_2$As$_2$ with indicated pressure as an implicit parameter, using
$T^*_{\rm NMR}$ extracted from NMR measurements (see Fig. \ref{fig:T1NMR}). Similarly, the red data shows
$T_{\rm c}$ vs $T^*$ for KFe$_2$As$_2$ using $T^*_{\rm R}$ extracted from resistivity measurements (see Fig. \ref{fig:rho}). 
}
\label{fig:Tstar}
\end{figure}

    We now consider the  empirical relation that $T_{\rm c}$ is proportional to $T^*$ observed in the AFe$_2$As$_2$ (A = K, Rb, Cs) family at ambient pressure \cite{Wu2016}. 
    Figure \ref{fig:Tstar} plots our results for $T_{\rm c}$ as a function of $T^*$ along with the results of Ref. \onlinecite{Wu2016}. 
    In the AFe$_2$As$_2$  (A= Cs, Rb, K) family at ambient pressure, $T_{\rm c}$ moves in proportion to $T^*$, suggesting that the change of $T^*$
is the primary factor in determining $T_{\rm c}$. 
     In contrast, for pressurized KFe$_2$As$_2$ we find that $T_{\rm c}$ decreases sharply as a function of $T^*$ below $p_{\rm c}\sim1.8$ GPa  and then becomes roughly independent of $T^*$ above $p_{\rm c}$.
These results indicate that $T^*$ is not the primary driver of the pressure dependence of $T_{\rm c}$ in KFe$_2$As$_2$. 
Instead, as will be described in the next section, we show the antiferromagnetic spin fluctuations play an important role for the pressure dependence of $T_{\rm c}$.
 
    Finally,  it is interesting to discuss the second empirical relation that $\gamma^{-1}\propto T^*$ under pressure.
  Quantum oscillation experiments under high pressure found that the effective mass $m^*$ decreases under pressure \cite{Terashima2014}.
     In addition, the coefficient $A$ in the low-temperature resistivity $\rho=\rho_0+AT^2$ decreases smoothly, which is also consistent with a decreasing $m^*$ under pressure \cite{Taufour2014}.
     The decreasing $m^*\sim\gamma$ accompanied by the increase of $T^*$ suggest that the $\gamma^{-1}\propto T^*$ relationship seems to hold under pressure, similar to the case of  AFe$_2$As$_2$ (A = K, Rb, Cs).
   As one moves from CsFe$_2$As$_2$ to RbFe$_2$As$_2$ to KFe$_2$As$_2$, the chemical pressure increases due to the decreasing size of the alkali metal ion \cite{Wang2016}. Simultaneously, $T^*$ increases \cite{Wu2016}. 
    Consequently, the increase of $T^*$ in KFe$_2$As$_2$ under physical pressure could be considered an extension of the chemical pressure trend. 
     However, it is noted that the  $\gamma^{-1}\propto T^*$ relationship does not appear to hold the case of carrier doping in  Ba$_{1-x}$K$_x$Fe$_2$As$_2$
     as seen in Ref. \onlinecite{Hardy2016}.

\subsection{D.  NQR Spin-Lattice Relaxation Rate}

\begin{figure}[tb]
\centering
\includegraphics[width=\columnwidth]{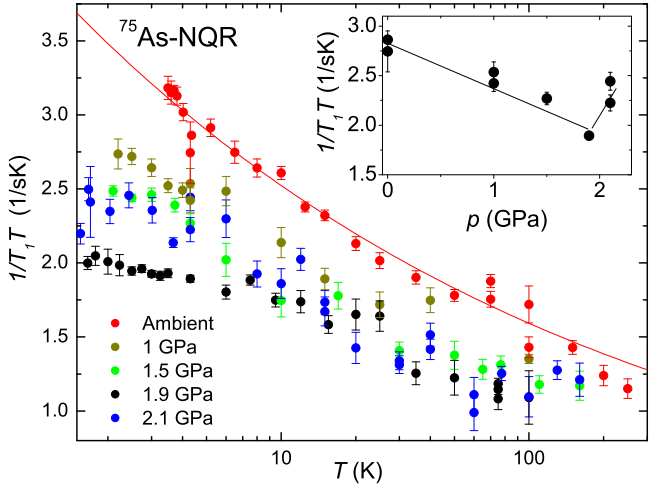}

\caption{
NQR $1/T_1T$ above $T_{\rm c}$ for various pressures. 
The solid red curve for ambient pressure is a power law fit (see text). 
Inset: The value of NQR $1/T_1T$ at 4.2 K as a function of pressure. 
}
\label{fig:T1NQR}
\end{figure}

Since $T^*$ evolves smoothly across the critical pressure $p_{\rm c}$, the pressure dependence of 
the coherence/incoherence crossover behavior cannot explain the non-monotonic behavior of $T_{\rm c}$ under pressure
in KFe$_2$As$_2$. To address this question, we have also performed NQR $1/T_1$ measurements
in both the PM and SC states. 
{No external magnetic field is required to measure NQR $1/T_1$, making this technique ideal for investigation of the SC state.

\subsubsection{1.  Paramagnetic State}
\label{NQRT1PM}
First, we consider the NQR $1/T_1T$ in the PM state at ambient pressure. 
As seen in Fig. \ref{fig:T1NQR}, the NQR $1/T_1T$ at ambient pressure follows a power law above $T_{\rm c}$: $1/T_1= 4T^{0.8}\Leftrightarrow1/T_1T= 4T^{-0.2}$ (shown by the red solid curve in Fig. \ref{fig:T1NQR}).
This power law is consistent with previously reported NQR results at ambient pressure \cite{Fukazawa2009}, and also NMR 1/$T_1$ data ($1/T_1 \propto T^{0.75}$)  described in the previous section.
In general, the nuclear spin-lattice relaxation rate measures the $\mathbf{q}$-summed dynamical susceptibility at the Larmor frequency
perpendicular to the quantization axis of nuclear spin, 
\begin{equation}
\frac{1}{T_1T}\sim\gamma^2k_B\sum_{\mathbf{q}}A_\perp^2(\mathbf{q})\frac{\rm{Im}\chi_\perp(\mathbf{q},\omega_L)}{\omega_L}.
\label{eq:T1}
\end{equation}
Therefore, since the NMR shift $K$, which reflects the $\mathbf{q}=0$ component of $\chi$, shows a weak temperature dependence,  the increase of $1/T_1T$ at low temperatures reflects the enhancement of low-energy $\mathbf{q}\neq0$ AFM spin fluctuations. 


   As shown in Fig. \ref{fig:T1NQR}, the enhancements of $1/T_1T$ at low temperatures seems to be suppressed up to $p_{\rm c}$ and then starts to increase above $p_{\rm c}$  with increasing pressure, although the pressure dependence of $1/T_1T$ becomes less clear at high temperatures above $\sim10$ K due to our experimental uncertainty.   
  To see clearly the pressure dependence of low temperature $1/T_1T$ data, we plot the  $1/T_1T$ values at 4.2 K as a function of pressure in the inset of Fig. \ref{fig:T1NQR}.
  Here we took the $1/T_1T$ values at 4.2 K because enhancements of $1/T_1T$ due to the AF spin fluctuations are more significant at low temperatures and also the temperature is close to the lowest temperature above  $T_{\rm c}$ in the paramagnetic state for all pressures measured. 
    The value of $1/T_1T$ at 4.2 K  clearly decreases with increasing pressure below $p_{\rm c}$ and  then increases again above $p_{\rm c}$.
    Since the value of  $1/T_1T$ reflects the strength of low-energy AFM spin fluctuations, we conclude that spin fluctuations at low temperatures are suppressed below $p_{\rm c}$ and enhanced again above $p_{\rm c}$. 
   This trend is very similar to the pressure dependence of $T_{\rm c}$.  
    Therefore, we may conclude that AFM spin fluctuations are involved in the superconducting pairing both above and below $p_{\rm c}$, consistent with the high-field NMR results \cite{Wang2016}.

\begin{figure}[tb]
\centering
\includegraphics[width=\columnwidth]{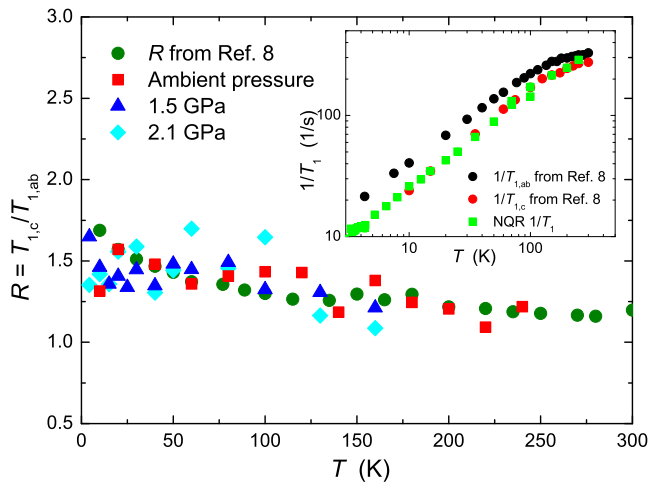}
\caption{
$T$ dependence of the ratio $R$ $\equiv$ $T_{\rm 1,c}$/$T_{\rm 1,ab}$ for different pressures where  NQR-$T_1$ data are used for $T_{\rm 1,c}$, in addition to $R$ obtained  from  $T_{\rm 1,c}$/$T_{\rm 1,ab}$ at ambient pressure from Ref. \onlinecite{Hirano2012}.  
NQR $1/T_1T$ above $T_{\rm c}$ for various pressures. 
The inset shows the $T$ dependence of 1/$T_1$ for $H||c$ and $H||ab$ from Ref. \onlinecite{Hirano2012},  together with NQR $1/T_1$ data.
}
\label{fig:Rratio}
\end{figure}

    However, it should be noted that the values of $1/T_1T$ decrease for the replacement of A from Cs to K in AFe$_2$As$_2$, despite the fact that $T_{\rm c}$  increases due to the replacement \cite {Wu2016}. 
   The relationship between $T_{\rm c}$ and 1/$T_1$ therefore appears to be different in the pressure and replacement cases. 
   Although at present the origin of the different behavior of $T_{\rm c}$ between the pressure and replacement cases is not understood well, we here discuss a few  possibilities to explain the difference. 

      One possible difference between the pressure and replacement cases may relate to the anisotropy of magnetic fluctuations. 
      According to Zhang ${\it et~al.}$ \cite{Zhang2017}, based on their NMR data,  
the anisotropy of the low-temperature AFM  fluctuations is found to significantly decrease with the replacement  from Cs to K in AFe$_2$As$_2$. 
      That is, the Cs sample with the lowest $T_{\rm c}$ in the family has the greatest anisotropy, suggesting that $T_{\rm c}$ may correlate with the anisotropy of the AFM fluctuations. 
     Zhang ${\it et~al}$. also suggested that the difference of the anisotropy may relate to  quantum criticality and that the Cs sample is the closest to a QCP. 

    It is interesting to compare this to the behavior of the magnetic fluctuation anisotropy in KFe$_2$As$_2$ under pressure which can be obtained by  taking a look at the ratio of 1/$T_1$ for the two field directions, $R$ $\equiv$ $(1/T_1)_{ab}/(1/T_1)_{c}$.
    According to the previous NMR studies performed on Fe pnictide SCs \cite{Kitagawa2009, KitagawaAFSF, Hirano2012}, the ratio $R$ depends on the nature of magnetic fluctuations and also anisotropy of the magnetic fluctuations as 
\begin{eqnarray}
R =  \left\{
\begin {array}{ll}

0.5 + \left(\frac{S_{ab}}{S_c}\right) ^2    \mbox{~ for  the stripe AFM fluctuations} \\ 

0.5     \mbox{~ for  the N\'eel-type spin fluctuations} \\
\end {array}
\right .
\label{eqn:correlation}
\end{eqnarray}
where  ${\cal S}_{\alpha}$ is the amplitude of the spin fluctuation spectral density at NMR frequency along the $\alpha$ direction. 
     Unfortunately, since we used a powder sample to improve the signal intensity, only $H||ab$ plane 1/$T_1$ NMR measurements are feasible. 
     Nevertheless, we can obtain some information about the anisotropy of the AFM spin fluctuations using our NQR 1/$T_1$ data. 
    Since the quantization axis of the electric field gradient is parallel to the $c$ axis, the NQR 1/$T_1$ should reflect magnetic fluctuations perpendicular to the $c$ axis. 
      These are the same fluctuations observed by NMR 1/$T_1$ for $H||c$ axis, where the quantization axis is determined by the magnetic field. 
    Indeed, we confirmed that our NQR 1/$T_1$ data coincide almost perfectly with the NMR 1/$T_1$ data under $H||c$ axis reported previously at ambient pressure \cite{Hirano2012}, as shown in the inset of Fig. \ref{fig:Rratio}. 
     This also indicates no magnetic field effects on 1/$T_1$.  
    Therefore, using both the NQR 1/$T_1$ and NMR 1/$T_1$ data under pressure, we can estimate how the anisotropy of magnetic fluctuations changes with pressure. 
    The estimated $R$ values using both the NQR 1/$T_1$ and NMR 1/$T_1$ data are shown in Fig. \ref{fig:Rratio} as a function of temperature for different pressures. 
    All $R$ values are greater than unity, consistent with the stripe-type spin fluctuations. 
    As shown, $R$ does not show any significant change with pressure. 
    This indicates that the anisotropy of spin fluctuations is almost independent of pressure, in contrast to the case of replacement effects on AFe$_2$As$_2$. 
     We suggest that the different behaviors of the spin fluctuation anisotropy between the pressure and replacement cases may be related to the different behavior of $T_{\rm c}$ in the two cases. 
    It is also interesting to note that several papers have proposed that, in the proximity of a QCP, the critical fluctuations may actually be detrimental to superconductivity in these systems \cite{Eilers2016, Hardy2013, Zhang2017}. 
   Since CsFe$_2$As$_2$ is considered  to be the closest to the QCP, it would be expected to have a low $T_{\rm c}$.

     It is also interesting to note in this context that in the hole-overdoped region of the Ba$_{1-x}$K$_x$Fe$_2$As$_2$ phase diagram, the AFM spin fluctuations and Sommerfeld coefficient determined by specific heat measurements are both enhanced with increasing $x$ while $T_{\rm c}$ decreases, similar to the case of AFe$_2$As$_2$ (A = K, Rb, Cs). 
    One possible explanation for the decrease of $T_{\rm c}$ in  Ba$_{1-x}$K$_x$Fe$_2$As$_2$ is the growth of competing ferromagnetic (FM) spin fluctuations, which coexist with the AFM spin fluctuations \cite{Wiecki2015}. 
   As demonstrated by Wiecki ${\it et~al}$., the growth of the AFM fluctuations with increasing $x$ in Ba$_{1-x}$K$_x$Fe$_2$As$_2$ is accompanied by the simultaneous growth of FM fluctuations. 
    These FM fluctuations may interfere with the AFM-fluctuation-based Cooper-pairing mechanism, thus lowering $T_{\rm c}$ despite the enhancement of AFM fluctuations. 
     It is possible such physics could apply to the AFe$_2$As$_2$ (A=K, Rb, Cs) system also.

\begin{figure}[tb]
\centering
\includegraphics[width=\columnwidth]{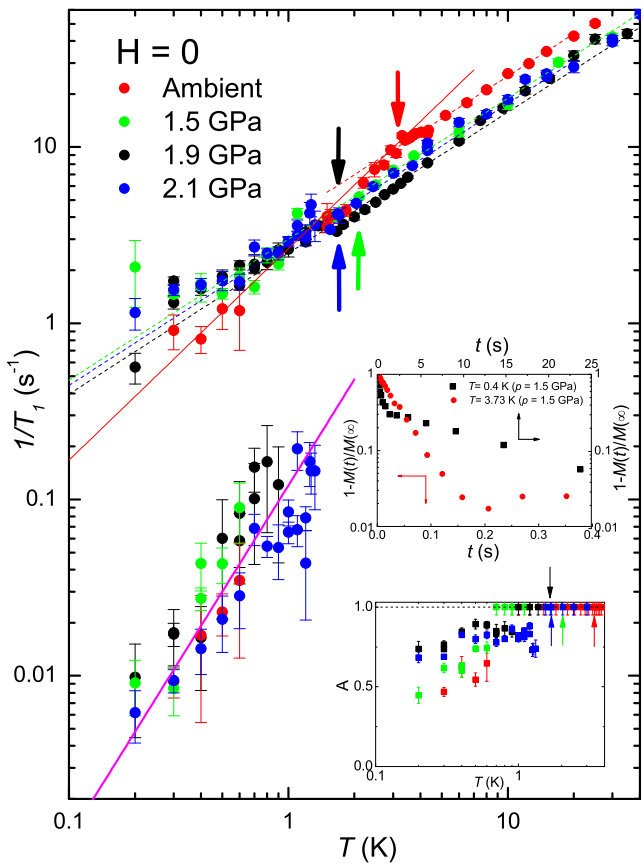}

\caption{NQR $1/T_1$ for indicated pressures. The arrows denote $T_{\rm c}$ at each pressure. The dashed lines are power law fits to the PM state data for each pressure.  The red solid line below $T_{\rm c}$ shows the power law with  $1/T_{\rm 1S}$ $\propto$ $T^{1.3}$ at ambient pressure.
Below $T\sim1$ K, a component with long $T_1$ appears.  
The solid pink line represents  $1/T_{\rm 1L}$ $\propto$ $T^2$ behavior. 
Upper inset: The typical two exponential behavior (black squares) of the nuclear magnetization recovery curve observed at low temperature ($T$ = 0.4 K; $p$ = 1.5 GPa), together with a single exponential behavior (red circles) at $T$ = 3.73 K and $p$ = 1.5 GPa.
Lower Inset: Fraction $A$ of nuclei relaxing with the short $T_1$ (see text). 
}
\label{fig:T1dilution}
\end{figure}

\subsubsection{2.  Superconducting State }

The $T$ dependence of NQR $1/T_1$ below $T_{\rm c}$ is shown in Fig. \ref{fig:T1dilution}.
At ambient pressure, $1/T_1$ follows the power law $1/T_1\sim T^{0.8}$ in the PM state as discussed above (red dashed line in Fig. \ref{fig:T1dilution}). 
A clear kink is seen at $T_{\rm c}$, and the data follow a new power law $1/T_1\sim T^{1.3}$ below $T_{\rm c}$ (red solid line in Fig. \ref{fig:T1dilution}). 
This behavior is consistent with previous ambient pressure NQR results \cite{Fukazawa2009}.
However, in contrast to Ref. \onlinecite{Fukazawa2009}, a long $T_1$ component is found to appear below $T=1$~K at ambient pressure and also under pressure.
The upper inset of Fig. \ref{fig:T1dilution} shows the typical two-component exponential behavior of the nuclear magnetization recovery curve observed at low temperature ($T$ = 0.4 K; $p$ = 1.5 GPa), together with a single exponential behavior at $T$ = 3.73 K and $p$ = 1.5 GPa.
Then,  we fit the recovery curves according to 
\begin{equation}
1-\frac{m(t)}{m(\infty)}=A\exp{(-3t/T_{\rm 1S})}+(1-A)\exp{(-3t/T_{\rm 1L})},
\end{equation}
where $T_{\rm 1S}$ and $T_{\rm 1L}$ are the short and long relaxation times, respectively. 
The parameter $A$, representing the fraction of nuclei relaxing with the shorter relaxation time $T_{\rm 1S}$, is shown in the lower inset of Fig. \ref{fig:T1dilution}, 
demonstrating that the long $T_{\rm 1L}$ component fraction increases with decreasing temperature.

    The existence of two $T_1$ components implies the existence of two distinct local electronic environments, which are physically separated in real space.
Similar two-component relaxation has been observed by NQR in the closely-related sample RbFe$_2$As$_2$, in which the two-component behavior was argued to be associated with a charge order of nanoscale periodicity \cite{Civardi2016}. 
    While we find no direct evidence for charge order in KFe$_2$As$_2$ in this study, charge ordering in KFe$_2$As$_2$ at 2.4 GPa 
(above our maximum pressure) was proposed by high pressure NMR \cite{Wang2016}. 
    Two-component relaxation has also been reported in CsFe$_2$As$_2$ under magnetic field in Refs. \onlinecite{Wu2016} (Supplemental Information), \onlinecite{Zhao2017} and \onlinecite{Li2016}.
   At present, although the origin of the two $T_1$ components in KFe$_2$As$_2$ is not clear, the similar behavior in closely related systems would suggest that the two-component behavior observed here is intrinsic. 
    Further studies will be needed to clarify the origin.  

    NQR $1/T_1$ is a sensitive probe of the reduction of the density of states (DOS) at the Fermi energy $N(E_{\rm F})$ due to the opening 
of the SC gap. In general, 1/$T_1$ in the SC state is given by \cite{Walstedt} 
\begin{equation}
\frac{1}{T_1}\sim\int_0^\infty\left[N^2_s(E)+M^2_s(E)\right]f(E)(1-f(E)){\rm{d}}E,
\label{eq:T1sc}
\end{equation}
where $N_s(E)$ is the DOS and $f(E)$ is the Fermi distribution function. $M_s(E)$ is the anomalous DOS
arising from Cooper pair coherence. Due to the lack of a coherence peak just below $T_{\rm c}$, we neglect the coherence term,
as has been done in previous NMR/NQR studies of FeAs superconductors. 

The very weak decrease of the short component $1/T_{\rm 1S}$ below $T_{\rm c}$ ($1/T_{\rm 1S}\sim T^{1.3}$), implies a very small SC gap.
Using a simple full gap model for $N_s(E)$, we estimate a gap of $\Delta(0)\sim0.07$ meV ($2\Delta(0)/k_{\rm B}T_{\rm c}\sim0.5$) from the short component, consistent with $2\Delta(0)/k_{\rm B}T_{\rm c}\sim0.51$ reported by previous NQR measurements \cite{Fukazawa2009}. 
For all but the lowest temperatures measured, the relaxation is dominated by the short component, as shown by the inset of Fig. \ref{fig:T1dilution}.
This implies that a large number fraction of nuclei see a nearly gapless electronic environment below $T_{\rm c}$.
This may correspond to a large ungapped DOS below $T_{\rm c}$ in KFe$_2$As$_2$ observed by scanning tunneling spectroscopy (STS) \cite{Fang2015}.
The large ungapped DOS was attributed to a Van Hove singularity just below the Fermi level seen by angle resolved photoemission spectroscopy (ARPES) \cite{Fang2015}.
   It is also worth mentioning that a residual DOS in SC state has been reported in SrFe$_2$As$_2$ under high pressure \cite{KitagawaSrFe2As2} and also in Co doped BaFe$_2$As$_2$ by specific heat measurements \cite{Hardy2010}. It is also suggested theoretically that the residual DOS is due to a possible formation of domain walls inherent to antiferromagnetism in iron pnicitde SCs \cite{Gorkov2010}.

     In contrast, the long component $1/T_{\rm 1L}$ shows a large reduction relative to the $1/T_1$ in the PM state, implying a large reduction
in $N_s(E_F)$ due to the SC gap. 
    Although the experimental uncertainty is large, 1/$T_{\rm 1L}$ seems to be proportional to $T^{2\pm1}$ as shown by the solid line in Fig. \ref{fig:T1dilution}.
   The sizable depletion of $N_s(E_{\rm F})$ only below $T\sim1$ K has been observed by the STS and ARPES experiments \cite{Fang2015}. 
The co-existence of one large gap and at least one very small gap has  also been reported with specific heat \cite{Hardy2014} and small angle neutron
scattering \cite{KawanoFurukawa2011} experiments. 
    However, from the two-component relaxation behavior, our NQR data suggest a real-space modulation of the local gap structure, which has not been reported previously.

      Under high pressure of 1.5 GPa and above, no obvious change of the slope of the short $T_1$ component occurs across $T_{\rm c}$ within our experimental uncertainty.
This indicates that the nuclei relaxing according to $1/T_{\rm 1S}$ see a gapless local electronic environment above 1.5 GPa.  
Therefore the small gap seen by $1/T_{\rm 1S}$ at ambient pressure is thought to be suppressed to zero near $p_{\rm c}$, and is not recovered above $p_{\rm c}$. 
Similarly, muon spin rotation ($\mu$SR) measurements \cite{Shermadini2012} on the closely-related RbFe$_2$As$_2$ with $p_{\rm c}\sim1.1$ GPa \cite{Tafti2015} reported that the smaller of two SC gaps is suppressed to zero near 1 GPa. 
    As for the long $T_1$ component under high pressure, as shown in Fig. \ref{fig:T1dilution}, no obvious change in 1/$T_{\rm 1L}$ can be found, suggesting no dramatic change in the magnitude of the larger SC gap upon pressure application.      
    According to Ref. \onlinecite{Taufour2014}, the SC gap structure changes above $p_{\rm c}$, where the SC gap is modulated along $k_z$. 
However, we did not observe a clear change in gap symmetry across $p_{\rm c}$ from our $1/T_1$ measurements.

\section{IV.  Conclusions}
We have presented $^{75}$As-NMR, NQR and resistivity data which clearly show an increase of the coherence/incoherence crossover temperature $T^*$ in KFe$_2$As$_2$ under pressure. 
   This increase of $T^*$ is expected due to the increase in hybridization between localized and conducting bands caused by pressure application. 
    We find that the relation $\gamma^{-1}\sim T^*$ observed in ambient pressure AFe$_2$As$_2$ (A = K, Rb, Cs) continues to hold under pressure.
However, the proportionality between $T^*$ and $T_{\rm c}$ is clearly broken under pressure. 
    The non-monotonic behavior of $T_{\rm c}$ under pressure is therefore unrelated to the coherence-incoherence crossover behavior in the paramagnetic state. However,  the strength of AFM spin fluctuations in the paramagnetic state is found to correlate with $T_{\rm c}$, evidencing clearly that the AFM spin fluctuations play an important role for the appearance of superconductivity in KFe$_2$As$_2$, although such a correlation cannot be seen in the replacement effects of A in the AFe$_2$As$_2$ (A= K, Rb, Cs) family.
    In the superconducting state, two $T_1$  components are observed at low temperatures, suggesting the existence of two distinct local electronic environments. 
    The temperature dependence of the short $T_{\rm 1s}$ indicates nearly gapless state below $T_{\rm c}$. 
    On the other hand, the temperature dependence of the long component 1/$T_{\rm 1L}$ implies a large reduction in the density of states at the Fermi level due to the SC gap formation.  
    These results suggest a real-space modulation of the local SC gap structure in KFe$_2$As$_2$ under pressure.


\section{V.  Acknowledgments}
     We thank M. Tanatar for helpful discussions. 
The research was supported by the U.S. Department of Energy, Office of Basic Energy Sciences, Division of Materials Sciences and Engineering. Ames Laboratory is operated for the U.S. Department of Energy by Iowa State University under Contract No.~DE-AC02-07CH11358.
Work at Argonne National Laboratory was supported by the U.S. Department of Energy, Office of Science, Basic Energy Sciences  under Contract No.~DE-AC02-06CH11357.


\begin{thebibliography}{99}
\bibitem{Canfield2010}
P. C. Canfield and S. L. Bud'ko, Annu. Rev. Condens. Matter Phys. {\bf 1}, 27 (2010).
\bibitem{Johnston2010} 
D. C. Johnston, Adv.  Phys. {\bf 59}, 803 (2010).
\bibitem{Fernandes2014}
R. M. Fernandes, A. V. Chubukov, and J. Schmalian, Nat. Phys. {\bf 10} 97 (2014).
\bibitem{Si2016}
Q. Si, R. Yu, and E. Abrahams, Nat. Rev. Mats. {\bf 1}, 16017 (2016).
doi:10.1038/natrevmats.2016.17 
\bibitem{Dai2015}
P. Dai, Rev. Mod. Phys. {\bf 87} 855 (2015).
\bibitem{Hardy2013}
F. Hardy, A. E. B\"{o}hmer, D. Aoki, P. Burger, T. Wolf, P. Schweiss, R. Heid, P. Adelmann, Y. X. Yao, G. Kotliar, J. Schmalian, and C. Meingast
Phys. Rev. Lett. {\bf 111}, 027002 (2013).
\bibitem{Wiecki2015}
P. Wiecki, B. Roy, D.~C. Johnston, S.~L. Bud'ko, P.~C. Canfield, and Y.~Furukawa,
Phys. Rev. Lett. {\bf 115}, 137001 (2015).
\bibitem{Hirano2012}
   M. Hirano, Y. Yamada, T. Saito, R. Nagashima, T. Konishi, T. Toriyama, Y. Ohta, H. Fukazawa, Y. Kohori, Y. Furukawa, K. Kihou, C.-H. Lee, A. Iyo, H. Eisaki,
J. Phys. Soc. Jpn., 81, 054704 (2012).
\bibitem{Wang2016}
P. S. Wang, P. Zhou, J. Dai, J. Zhang, X. X. Ding, H. Lin, H. H. Wen, B. Normand, R. Yu, and W. Yu, Phys. Rev. B {\bf 93}, 085129 (2016).
\bibitem{Ding2008}
H. Ding, P. Richard, K. Nakayama, K. Sugawara, T. Arakane, Y. Sekiba, A. Takayama, S. Souma, T. Sato, T. Takahashi, Z. Wang, X. Dai, Z. Fang, G. F. Chen, J. L. Luo, and N. L. Wang, Europhys. Lett. {\bf 83} 47001 (2008).
\bibitem{Fukazawa2009}
H. Fukazawa, Y. Yamada, K. Kondo, T. Saito, Y. Kohori, K. Kuga, Y. Matsumoto, S. Nakatsuji, H. Kito, P. M. Shirage, K. Kihou, N. Takeshita, C.-H. Lee, A. Iyo, and H. Eisaki, J. Phys. Soc. Jpn. {\bf 78}, 083712 (2009).
\bibitem{Hashimoto2010}
K. Hashimoto, A. Serafin, S. Tonegawa, R. Katsumata, R. Okazaki, T. Saito, H. Fukazawa, Y. Kohori, K. Kihou, C. H. Lee, A. Iyo, H. Eisaki, H. Ikeda, Y. Matsuda, A. Carrington, and T. Shibauchi, Phys. Rev. B {\bf 82}, 014526 (2010).
\bibitem{KawanoFurukawa2011}
H. Kawano-Furukawa, C. J. Bowell, J. S. White, R. W. Heslop, A. S. Cameron, E. M. Forgan, K. Kihou, C. H. Lee, A. Iyo, H. Eisaki, T. Saito, H. Fukazawa, Y. Kohori, R. Cubitt, C. D. Dewhurst, J. L. Gavilano, and M. Zolliker, Phys. Rev. B {\bf 84}, 024507 (2011).
\bibitem{Reid2012}
J. P. Reid, M. A. Tanatar, A. Juneau-Fecteau, R. T. Gordon, S. R. de Cotret, N. Doiron-Leyraud, T. Saito, H. Fukazawa, Y. Kohori, K. Kihou, C. H. Lee, A. Iyo, H. Eisaki, R. Prozorov, and L. Taillefer, Phys. Rev. Lett. {\bf 109}, 087001 (2012).
\bibitem{Okazaki2012}
K. Okazaki, Y. Ota, Y. Kotani, W. Malaeb, Y. Ishida,  T. Shimojima, T. Kiss, S. Watanabe, C. T. Chen, K. Kihou, C. H. Lee, A. Iyo, H. Eisaki, T. Saito, H. Fukazawa, Y. Kohori, K. Hashimoto, T. Shibauchi, Y. Matsuda, H. Ikeda, H. Miyahara, R. Arita, A. Chainani, and S. Shin, Science {\bf 337}, 1314 (2012).
\bibitem{Ota2014}
Y. Ota, K. Okazaki, Y. Kotani, T. Shimojima, W. Malaeb, S. Watanabe, C.-T. Chen, K. Kihou, C. H. Lee, A. Iyo, H. Eisaki, T. Saito, H. Fukazawa, Y. Kohori, and S. Shin
Phys. Rev. B {\bf 89}, 081103(R) (2014).
\bibitem{Hardy2014}
F. Hardy, R. Eder, M. Jackson, D. Aoki, C. Paulsen, T. Wolf, P. Burger, A. B\"{o}hmer, P. Schweiss, P. Adelmann, R. A. Fisher, and C. Meingast,
J. Phys. Soc. Jpn. {\bf 83} 014711 (2014).
\bibitem{Tafti2013}
F. F. Tafti, A. Juneau-Fecteau, M.-E. Delage, S. R. de Cotret, J.-P. Reid, A. F. Wang, X.-G. Luo, X. H. Chen, N. Doiron-Leyraud, and L. Taillefer, Nat. Phys. {\bf 9}, 349 (2013).
\bibitem{Terashima2014}
T. Terashima, K. Kihou, K. Sugii, N. Kikugawa, T. Matsumoto, S. Ishida, C.-H. Lee, A. Iyo, H. Eisaki, and S. Uji, Phys. Rev. B {\bf 89}, 134520 (2014).
\bibitem{Taufour2014}
V. Taufour, N. Foroozani, M. A. Tanatar, J. Lim, U. Kaluarachchi, S. K. Kim, Y. Liu, T. A. Lograsso, V. G. Kogan, R. Prozorov, S. L. Bud'ko, J. S. Schilling, and P. C. Canfield
Phys. Rev. B {\bf 89}, 220509(R) (2014).
\bibitem{Hong2013}
X. C. Hong, X. L. Li, B. Y. Pan, L. P. He, A. F. Wang, X. G. Luo, X. H. Chen, and S. Y. Li, Phys. Rev. B {\bf 87}, 144502 (2013).
\bibitem{Tafti2014}
F. F. Tafti, J. P. Clancy, M. Lapointe-Major, C. Collignon, S. Faucher, J. A. Sears, A. Juneau-Fecteau, N. Doiron-Leyraud, A. F. Wang, X.-G. Luo, X. H. Chen, S. Desgreniers, Y.-J. Kim, and L. Taillefer, Phys. Rev. B {\bf 89}, 134502 (2014).
\bibitem{Zhang2015}
Z. Zhang, A. F. Wang, X. C. Hong, J. Zhang, B. Y. Pan, J. Pan, Y. Xu, X. G. Luo, X. H. Chen, and S. Y. Li, Phys. Rev. B {\bf 91}, 024502 (2015).
\bibitem{Tafti2015}
F. F. Tafti, A. Ouellet, A. Juneau-Fecteau, S. Faucher, M. Lapointe-Major, N. Doiron-Leyraud, A. F. Wang, X.-G. Luo, X. H. Chen, and L. Taillefer, Phys. Rev. B {\bf 91}, 054511 (2015).
\bibitem{Mizukami2016}
Y. Mizukami, Y. Kawamoto, Y. Shimoyama, S. Kurata, H. Ikeda, T. Wolf, D. A. Zocco, K. Grube, H. v. L\"ohneysen, Y. Matsuda, and T. Shibauchi
Phys. Rev. B {\bf 94}, 024508 (2016).
\bibitem{Eilers2016}
F. Eilers, K. Grube, D. A. Zocco, T. Wolf, M. Merz, P. Schweiss, R. Heid, R. Eder, R. Yu, J.-X. Zhu, Q. Si, T. Shibauchi, and H. v. L\"{o}hneysen, Phys. Rev. Lett. {\bf 116}, 237003 (2016).
\bibitem{Civardi2016}
E. Civardi, M. Moroni, M. Babij, Z. Bukowski, and P. Carretta, Phys. Rev. Lett. {\bf 117}, 217001 (2016).
\bibitem{Wu2016}
   Y.~P. Wu, D. Zhao, A.~F. Wang, N.~Z. Wang, Z.~J. Xiang, X.~G. Luo, T. Wu, and X.~H. Chen, Phys. Rev. Lett. {\bf 116}, 147001 (2016).
\bibitem{Zhao2017}
D. Zhao, S. J. Li, N. Z. Wang, J. Li, D. W. Song, L. X.Zheng, L. P. Nie, X. G. Luo, T. Wu, X. H. Chen,
arXiv:1705.09885 [cond-mat.supr-con].
\bibitem{Gegenwart2008}
P. Gegenwart, Q. Si, F. Steglich, Nat. Phys. {\bf 4} 186 (2008). 
\bibitem{Curro2009}
N. J. Curro, Rep. Prog. Phys. {\bf 72} 026502 (2009).

\bibitem{You2011} Y.-Z. You, F. Yang, S.-P. Kou, and Z.-Y. Weng, Phys. Rev. B {\bf 84}, 054527 (2011).
\bibitem{Gorkov2013} L. P. Gor'kov and G. B. Teitel'baum, Phys. Rev. B {\bf 87}, 024504 (2013).


\bibitem{Fang2015}
D. Fang, X. Shi, Z. Du, P. Richard, H. Yang, X. X. Wu, P. Zhang, T. Qian, X. Ding, Z. Wang, T. K. Kim, M. Hoesch, A. Wang, X. Chen, J. Hu, H. Ding, and H.-H. Wen,  Phys. Rev. B {\bf 92}, 144513 (2015).
\bibitem{Georges2013}
A. Georges, L. de' Medici, and J. Mravlje, Annu. Rev. Condens. Matter Phys. {\bf 4} 137 (2013). 
\bibitem{Hardy2016}
F. Hardy, A. E. B\"{o}hmer, L. de' Medici, M. Capone, G. Giovannetti, R. Eder, L. Wang, M. He, T. Wolf, P. Schweiss, R. Heid, A. Herbig, P. Adelmann, R. A. Fisher, and C. Meingast, Phys. Rev. B {\bf 94}, 205113 (2016).
\bibitem{Pines2013}
D. Pines, J. Phys. Chem. B 117, 13145 (2013).
\bibitem{Fujiwara2007}
   N. Fujiwara, T. Matsumoto, K. Koyama-Nakazawa, A. Hisada and Y. Uwatoko, Rev. Sci. Instrum. {\bf 78}, 073905 (2007).
\bibitem{Aso2007}
   N. Aso, T. Fujiwara, Y. Uwatoko, H. Miyano, H. Yoshizawa, J. Phys. Soc. Jpn. {\bf 76} Suppl. A, pp. 228-229 (2007). 
\bibitem{Fukazawa2007}
    H. Fukazawa, N. Yamatoji, Y. Kohori, C. Terakura, N. Takeshita, Y. Tokura and H. Takagi, Rev. Sci. Instrum. {\bf 78}, 015106 (2007).
\bibitem{Reyes1992}
    A. P. Reyes, E. T. Ahrens, R. H. Heffner, P. C. Hammel, and J. D. Thompson, Rev. Sci. Instrum. {\bf 63}, 3120 (1992).
\bibitem{Colombier2007}
E. Colombier and D. Braithwaite, Rev. Sci. Instrum. {\bf 78}, 093903 (2007).
\bibitem{Grinenko2014}
V. Grinenko, W. Schottenhamel, A. U. B. Wolter, D. V. Efremov, S.-L. Drechsler, S. Aswartham, M. Kumar, S. Wurmehl, M. Roslova, I. V. Morozov, B. Holzapfel, B. B\"{u}chner, E. Ahrens, S. I. Troyanov, S. K\"{o}hler, E. Gati, S. Kn\"{o}ner, N. H. Hoang, M. Lang, F. Ricci, and G. Profeta,
Phys. Rev. B {\bf 90}, 094511 (2014).
\bibitem{Budko2012}
S. L. Bud'ko, Y. Liu, T. A. Lograsso, and P. C. Canfield
Phys. Rev. B {\bf 86}, 224514 (2012).
\bibitem{Slichter_book} C. P. Slichter, {\it Principles of Magnetic Resonance}, 3rd ed. (Springer, New York, 1990).
\bibitem{Mossbaure} 
K. Nishiyama, F. Dimmling, Th. Kornrumpf, and D. Riegel,
Phys. Rev. Lett. {\bf 37} 357 (1976).
\bibitem{Budko2017}
S. L. Bud'ko, T. Kong , W. R. Meier, X. Ma and P. C. Canfield, Philosophical Magazine, {\bf 97}, 2689, (2017).
\bibitem{Cui2017} J. Cui, Q.-P. Ding, W. R. Meier, A. E. B\"ohmer, T. Kong, V. Borisov, Y. Lee, S. L. Bud'ko, R. Valent\'i, P. C. Canfield, and Y. Furukawa, Phys. Rev. B {\bf 96}, 104512 (2017).
\bibitem{Kohori2001}
Y. Kohori, Y. Yamato, Y. Iwamoto, T. Kohara, E. D. Bauer, M. B. Maple, and J. L. Sarrao, Phys. Rev. B {\bf 64}, 134526 (2001).
\bibitem{Kohara1986}
T. Kohara, Y. Kohori, K. Asayama, Y. Kitaoka, M.B. Maple, and M. S. Torikachvili, Solid State Commun. {\bf 59}, 603 (1986).
\bibitem{Yang2008} Y.-F. Yang, Z. Fisk, H.-O. Lee, J.D. Thompson, and D. Pines, Nature (London) {\bf 454}, 611 (2008).

\bibitem{Zhang2017} Z. T. Zhang, D. Dmytriieva, S. Molatta, J. Wosnitza, S. Khim, S. Gass, A.U.B. Wolter, S. Wurmehl, H.-J. Grafe, H. K\"{u}hne, arXiv:1703.00780. 
\bibitem{Kitagawa2009} K. Kitagawa, N. Katayama, K. Ohgushi, and M. Takigawa, J. Phys. Soc. Jpn. {\bf 78}, 063706 (2009).
\bibitem {KitagawaAFSF} S. Kitagawa, Y. Nakai, T. Iye, K. Ishida, Y. Kamihara, M. Hirano, and H. Hosono, Phys. Rev. B {\bf 81}, 212502 (2010). 


\bibitem{Li2016}
J. Li, D. Zhao, Y. P. Wu, S. J. Li, D. W. Song, L. X. Zheng, N. Z. Wang, X. G. Luo, Z. Sun, T. Wu, X. H. Chen,
arXiv:1611.04694. 
\bibitem{Walstedt}
   R. E. Walstedt, {\it The NMR Probe of High-$T_{\rm c}$ Materials}, STMP 228 (Springer, Berlin Heidelberg 2008).
\bibitem{KitagawaSrFe2As2}K. Kitagawa, N. Katayama, H. Gotou, T. Yagi, K. Ohgushi, T. Matsumoto, Y. Uwatoko, and M. Takigawa, Phys. Rev. Lett.  {\bf 103}, 257002 (2009).
\bibitem{Hardy2010} F. Hardy, T. Wolf, R. A. Fisher, R. Eder, P. Schweiss, P. Adelmann, H. v. Löhneysen, and C. Meingast, Phys. Rev. B {\bf 81}, 060501(R) (2010).
\bibitem{Gorkov2010} L. P. Gor'kov and G. B. Teitel'baum, Phys. Rev. B {\bf 82}, 020510(R) (2010).
\bibitem{Shermadini2012}
Z. Shermadini, H. Luetkens, A. Maisuradze, R. Khasanov, Z. Bukowski, H.-H. Klauss, and A. Amato, Phys. Rev. B {\bf 86} 174516 (2012).


\end{thebibliography}
\end{document}